\documentstyle[12pt]{article}

\newcommand{\eq}[1]{Eq.(\ref{#1})}
\def\a{\alpha}
\def\D{\Delta}
\def\be{\begin{equation}}
\def\ee{\end{equation}}
\begin{document}
\vspace{0.35in}
\begin{center}

{\Large Variational bound for the energy of two- dimensional \\
quantum antiferromagnet}
\vspace{0.33in}

{\large A.A.~Ovchinnikov} \\
\vspace{0.15in}

{\it Institute for Nuclear Research of the Russian Academy of Sciences,
\\ Moscow 117312, Russia} \\
\vspace{0.4in}

\end{center}
\begin{abstract}

We obtain the variational upper bound for the ground- state energy of
two-dimensional antiferromagnetic Heisenberg model on a square
lattice at arbitrary value of the anisotropy parameter using the
two-dimensional generalization of Jordan-Wigner transformation.
Our result can be considered as an
upper bound for the perturbation theory series about the Ising
limit.

\end{abstract}

At present time two dimensional quantum spin systems attract much
attention in connection with the problem of high-$T_c$
superconductivity. For the antiferromagnetic  Heisenberg model at
some values of the anisotropy parameter the existence of the
long-range order was proved \cite{L}, however the exact ground state
is not known. Apart from the linear spin wave theory \cite{A}
various methods to evaluate the ground state energy for
the Heisenberg antiferromagnet were proposed. For instance the
perturbation theory and the cluster expansion about the Ising limit
were used \cite{H}. However, although the convergence of the series
of the perturbation theory is  good these estimates are not the
variational ones. At the same time the energy corresponding to any
reasonable variational ground-state wave function cannot be computed
exactly (for example of these calculations see ref.\cite{S}).
Finally at present time the accuracy of the numerical simulations
\cite{NUM} is not sufficiently high.   In this context the variational
estimates of the ground state energy for the two-dimensional
antiferromagnetic Heisenberg model are of interest.

In the present letter we obtain the exact upper bound for the
ground-state energy of the $s=1/2$ quantum antiferromagnet for
arbitrary value of the anisotropy parameter. Our variational estimates
are sufficiently low and may be useful in connection with the study of
two-dimensional spin models in the framework of the other methods.

Our method is based on the transformation which change the statistics
of particles on a two- dimensional lattice.
There are several ways to define the Hamiltonian of
particles obeying the fractional statistics (anyons) on a
lattice (for example see ref.\cite{F}). We can use the most natural
form of the definition of anyon operators in terms of the fermions
\be
b_i^{+}(\a)=a_i^{+}\exp\left(-i\a\sum_{l\neq i}\phi_{il}n_l\right),
\hspace{0.2in} n_l=a_l^{+}a_l
\label{1}
\ee
where the operators $a_i^{+}, a_i$ obeys Fermi statistics and
$\phi_{il}$ is the angle between the direction from the site $i$ to
the site $l$ and some fixed direction, $x$- axis for example.
In accordance with the multi-valuedness of the anyon wave
function the operator $b_i^{+}(\a)$ is multi-valued at arbitrary
fractional value of the statistical parameter $\a$, which describes in
\eq{1} the deviation from the Fermi statistics. In particular, at
$\a=1$ the operators (\ref{1}) are the hard core boson operators,
which commute at different sites and behave like the fermions at the
same site. Expressing the spin operators ($s=1/2$) in terms of the
Holstein - Primakoff bose operators
\[
S_i^{+}=b_i^{+}, \hspace{0.2in} S_i^{-}=b_i, \hspace{0.2in}
S_i^z=b_i^{+}b_i-\frac{1}{2},
\]
we obtain the representation of spin operators in terms of the
fermions which can be thought of as a two dimensional generalization
of the well known Jordan - Wigner transformation for one dimension:
\[
S_i^{+}=a_i^{+}\exp\left(-i\sum_{l\neq i}\phi_{il}n_l\right),
\hspace{0.2in} S_i^z=a_i^{+}a_i-\frac{1}{2}.
\]
The Hamiltonian of the Heisenberg antiferromagnet
$H=\sum_{<ij>} \vec{S}_{i} \vec{S}_{j}$
has a complicated form
\be
H=-\frac{1}{2}\sum_{<ij>}\left( a_i^{+}a_j
\exp\left(-i\sum_{l\neq i,j}\phi_{ijl}n_l\right)+ h.c. \right)
+ \sum_{<ij>}(n_i -{1\over 2})(n_j -{1\over 2}),
\label{H0}
\ee
where $\phi_{ijl}=\phi_{il}-\phi_{jl}$ and $<ij>$ denotes the nearest
neighbour sites. The minus sign before the first term in \eq{H0}
is due to the redefinition of the operators
$a_i\rightarrow-a_i$ on one of the sublattices of the square lattice.
In order to simplify the Hamiltonian one can make the substitution
$n_l\rightarrow\bar{n}_l$ in the exponential of \eq{H0}, where
$\bar{n}_i$ is the average particle number at a given site,
$\bar{n}=1/2$ for the half filling ($S^z=0$),
which will be considered in the present paper.
This procedure is usually refered to as a (vector) mean field
(MF) approximation.  After this substitution the MF Hamiltonian
\be
H_{MF}=-\frac{1}{2}\sum_{<ij>}(\chi_{ij}a_i^{+} a_j+ h.c.)
+ U \sum_{<ij>}(n_i -{1\over 2})(n_j -{1\over 2}),
\label{MF}
\ee
where $\chi_{ij}=\exp(-i\sum_l\phi_{ijl}\bar{n}_l)$,
describes the system of fermions in the
homogeneous statistical magnetic field with the magnitude
corresponding to the flux $\phi=\pi$ through the plaquette.
The parameter $U=1$ for the isotropic model. Due to
the gauge invariance ($a_i\rightarrow a_i\exp(\theta_i)$) the phases
of $\chi_{ij}$ are depend on the
gauge fixing condition. The sum of the phases around the closed
contour is fixed and equal to the one-half flux quantum through the
plaquette for the case of the half filling. The eigenstates of the MF
Hamiltonian does not depend on the choice of the gauge.
The second term in \eq{MF} is the interaction of fermions.

We use the variational theorem proved in ref.\cite{C} for the hard
core bosons in the absence of the interaction term. Let
$\psi_{MF}(i_1,\ldots i_N)$ and $E_{MF}$ to be respectively the exact
ground-state wave function and the ground state energy
of the MF Hamiltonian
(\ref{MF}) ($i_1,\ldots i_N$ - are the particle positions,
$<\psi_{MF}|\psi_{MF}>=1$). Consider the contribution of a given bond
to the expectation value of \eq{MF} over the ground state. We have the
following inequality:
\[
-\sum_{i_2\ldots i_N}
|\psi_{MF}(i,i_2,\ldots i_N)| |\psi_{MF}(j,i_2,\ldots i_N)|
\]
\[
\leq ~~  -Re \left( \sum_{i_2\ldots i_N}
\psi_{MF}^{*}(i,i_2,\ldots i_N)\psi_{MF}(j,i_2,\ldots i_N)\right).
\]
The left-hand side of this inequality is the contribution to the
expectation value of the exact Hamiltonian (\ref{H0}) in the bosonic
representation. The normalization as well as the expectation value of
the operator given by the last term of \eq{MF} are the same for the
wave functions $\Psi_{MF}$ and $|\Psi_{MF}|$.
Thus it is proved that the ground- state energy of the
initial bosonic Hamiltonian $E_0$ is bounded from above by $E_{MF}$:
\[
E_{0} \leq  E_{MF}.
\]
This relation allows one to obtain an upper bound for the energy of
the antiferromagnet. We have to obtain the appropriate $variational$
estimate for the ground- state energy of the MF Hamiltonian
(\ref{MF}).
As a variational wave function let us choose the wave
function corresponding to the Hamiltonian which is obtained from
$H_{MF}$ in the mean field approximation in respect to the fermion
interaction. We assume the existence of Neel order in this state.
Linearising the interaction and using the substitution
$<n_i>\rightarrow (-1)^i \D/4$ (we use the notation
$(-1)^i=(-1)^{i_x+i_y}$) we obtain the
Hamiltonian
\be
-\frac{1}{2}\sum_{<ij>}(\chi_{ij}a_i^{+} a_j+ h.c.)
- \D \sum_{i}(-1)^i n_i.
\label{vmf}
\ee
In this formula $\D$ is the variational parameter which is to be
determined from the condition of minimum of the expectation value of
$H_{MF}$ in the state given by the ground state of the mean
field Hamiltonian \eq{vmf}.
This expectation value is the variational bound for the energy
$E_{MF}$. Note that the choice of the wave function
is consistent with the MF treatment of the statistical
interaction since the sum of the phases around the plaquette for the
Neel ordered state is the same as in the case of $\bar{n}_i=1/2$.

The calculations are most easily performed using the symmetric gauge
\[
\chi_{i,i+\hat{x}}=\frac{1}{\sqrt{2}}(1+i(-1)^i),~~~
\chi_{i,i+\hat{y}}=\frac{1}{\sqrt{2}}(1-i(-1)^i),
\]
where $\hat{x},\hat{y}$ are the unit vectors corresponding to the
lattice spacing. In the momentum space in terms of the doublets
$\psi_{1k}=(a_k,a_{k-Q})$
($0<k_x,k_y<\pi,~Q=(\pi,\pi)$) and
$\psi_{2k}=(a_k \\ a_{k-Q_1})$
($0<-k_x,k_y<\pi,~Q_{1}=(-\pi,\pi)$)
the equation (\ref{vmf}) has the form
\[
H = \sum_{k_x>0,~k_y>0}\psi_{1k}^{+}M_k \psi_{1k} +
    \sum_{k_x<0,~k_y>0}\psi_{2k}^{+}M_k \psi_{2k},
\]
where the matrix $M_k$ is
\[
M_k = - \left(
\begin{array}{cc}
   c_1   &  \D-ic_2 \\
\D+ic_2  &   -c_1
\end{array}  \right),
{}~~~            c_{1,2}={1\over\sqrt{2}}(\cos k_x \pm \cos k_y).
\]
The eigenvalues are $E_k=\pm(\cos^{2}k_x+\cos^{2}k_y+\D^2)^{1/2}$
where the momentum $k$ is restricted to the half of the Broullien
zone $k_y>0$. The negative energy levels are filled.
Let us calculate the average of $H_{MF}$ over this state. The values
of $<\chi_{ij}a_{i}^{+}a_{j}>$ for a given $<ij>$
does not depend on the choice of the gauge.
This values are real (and positive) which can
be deduced from the parity invariance of our state. The expectation
value of the second term in \eq{MF} is
$<n_{i}n_{j}>=<n_i><n_j>-<a_{i}^{+}a_j><a_{j}^{+}a_i>$. The expression
for the particle number at a given site has the
form $<n_i>=1/2+(-1)^i\D_{1}/4$, where the
parameter $\D_1$ does not coincide with the parameter $\D$. We obtain
for $\D_1$ and $\xi=|<a_{i}^{+}a_{j}>|$ (which is the same for all
bonds) the following expressions:
\[
\D_1 = 8 \sum_{k_x>0,~k_y>0} \frac{\D}{E_k},~~~~
\xi = \sum_{k_x>0,~k_y>0}
\frac{\cos^{2}k_x+\cos^{2}k_y}{E_k}.
\]
The final expression for the variational estimate is
\be
E^{var}/2L^2 = \frac{\D^2}{16U}- \sum_{k_x>0,~k_y>0} E_k
-\frac{U}{16}\left(\D_1-\frac{\D}{U}\right)^2
- \xi^2,
\label{var}
\ee
where $L^2$ is the number of sites.
The sum of the first two terms is the energy in
the mean field approximation \eq{vmf}.

At $U=0$ we get the exact energy of the MF Hamiltonian and the
corresponding
estimate for the energy for the XY model is $-0.2395$ per bond. In
comparison with the energy determined with the help of the numerical
simulations $-0.27(\pm 10\%)$ \cite{NUM}, the bound is too high.
It is less restrictive than the bound based on
the simple trial variational wave function wave function.
For example the energy corresponding to the Neel ordered state (in
the $y$-direction) is $-0.25$.
That is in agreement with the statement of ref's.\cite{C,O} that the
corrections due to the fluctuations around the average magnetic field
background are of order of unity. The situation is different for the
isotropic (XXX) model ($U=1$). In this case the perturbation theory
series is rapidly converges and the corrections due to the statistics
of particles are suppressed. For instance for the Hamiltonian
(\ref{H0}) the corrections to the MF approximation are of order
$\sim 1/(2U)^6$ which is a sufficiently small value \cite{O}.
In this sense our result can be considered as an
estimate from above for the perturbation theory series about
the Ising limit. It is difficult to establish the restrictions of this
type using the other methods.
Minimizing the expression (\ref{var}) with respect to $\D$
($\D_{0}=1.19$) we obtain
$E^{var}_{xxx}/2L^{2}=-0.33034$,
which is sufficiently good upper bound for the energy. For comparison
the best estimate obtained using the method of ref.\cite{H} is
$-0.334$. Note that although the prediction of the linear spin wave
theory \cite{A} is $-0.329$, this method does not result in the
correct ground-state wave function and this value cannot be considered
as a variational bound.

For the anisotropic model we proceed as follows. For simplicity
let us consider the axially symmetric model although our method can be
easily generalized to the case of arbitrary asymmetry.
We use the description in terms of the Holstein-Primakoff bosons for
the equivalent Hamiltonian
$H=\sum_{<ij>}(S_{i}^{x}S_{j}^{x}+\gamma S_{i}^{y}S_{j}^{y}+
S_{i}^{z}S_{j}^{z} )$.
After the substitution $b_i\rightarrow (-1)^{i}b_i$ we get
\be
H=-\sum_{<ij>}\left( \frac{1+\gamma}{4}(b_{i}^{+}b_j + h.c) +
\frac{1-\gamma}{4}(b_{i}^{+}b_{j}^{+} + h.c) \right)
+ \sum_{<ij>}(n_i-{1\over 2})(n_j-{1\over 2}).
\label{xy}
\ee
Consider the trial variational wave function with the fixed number of
bosons. For this state the expectation value of the second term
$\sim (b_{i}^{+}b_{j}^{+}+b_{i}b_{j})$ in \eq{xy} is zero. The
Hamiltonian
(\ref{xy}) without this term can be used to obtain the variational
estimate for the ground-state energy for arbitrary $\gamma$ according
to our method. Note that the contribution of the omitted term is small
in the framework of the perturbation theory \cite{H} since it appears
only in the fourth order.
The analysis can be performed at arbitrary value of the parameter
$\gamma$.
For the XY model ($\gamma=0$) which is
equivalent to the system of the hard core bosons at the half filling
we found the estimate $-0.26776$ per bond ($\D_{0}=3.4$).
This estimate is in agreement with result of the numerical simulations
\cite{NUM}.

In conclusion, although the wave function corresponding to the mean
field Hamiltonian (\ref{MF}) cannot be used to describe the long-range
properties of the model (for example, the energy of the low-lying
excitations) the ground-state energy can be estimated with the
sufficiently high accuracy. We found the $variational$ upper bound for
the ground- state energy of the two-dimensional Heisenberg
antiferromagnet on a square lattice at arbitrary value of the
anisotropy parameter. Our results can be thought of as a peculiar
upper bound for the perturbation theory series about the Ising limit
and may be useful in the context of the other approaches.

This work was supported, in part, by a Soros Foundation Grant
awarded by the American Physical Society.


\begin{thebibliography}{99}

\bibitem{L}
T.Kennedy, E.H.Lieb, B.S.Shastry, J.Stat.Phys. 53 (1988) 1019.

\bibitem{A}
P.W.Anderson, Phys.Rev. 86 (1952) 694.

\bibitem{H}
R.B.Pearson, Phys.Rev.B 16 (1977) 1109;
D.A.Huse, Phys.Rev.B 37 (1988) 2380;
C.J.Hamer, J.Oitmaa, Z.Weihong, Phys.Rev.B 43 (1991) 10789.

\bibitem{S}
S.Sachdev, Phys.Rev.B 39 (1988) 12232.

\bibitem{NUM}
J.Oitmaa, D.D.Betts, Can.J.Phys. 56 (1978) 897;
J.Oitmaa, D.D.Betts, L.G.Marland, Phys.Lett. 79A (1980) 193;
Y.Okabe, M.Kikuchi, J.Phys.Soc.Jpn. 57 (1988) 4351.

\bibitem{F}
E.Fradkin, Phys.Rev.Lett. 63 (1989) 322;\\
J.Frohlich, P.A.Marchetti, Commun.Math.Phys. 121 (1989) 177.

\bibitem{C}
C.Gros, S.M.Girvin, G.S.Canright, M.D.Johnson, Phys.Rev.B 43 (1991)
5883;
G.S.Canright, S.M.Girvin, A.Brass, Phys.Rev.Lett. 63 (1989) 2291.

\bibitem{O}
A.A. Ovchinnikov, Mod.Phys.Lett.B 6 (1992) 1951.



\end{thebibliography}
\end{document}